\documentclass[review]{elsarticle}

\usepackage{array}

\usepackage[symbol]{footmisc}

\journal{Journal of \LaTeX\ Templates}

\begin{document}

\begin{frontmatter}

\title{Quenching factor measurement for  NaI(Tl) scintillation crystal}

\author{H.W. Joo\textsuperscript{1,2}, 
H.S. Park\textsuperscript{1,}\footnote[2]{Corresponding author. E-mail : hyeonseo@kriss.re.kr}, 
J.H. Kim\textsuperscript{1}, 
J.Y. Lee\textsuperscript{1}, 
S.K. Kim\textsuperscript{2,}\footnote[1]{Corresponding author. E-mail : skkim@snu.ac.kr}, 
Y.D. Kim\textsuperscript{3}, H.S. Lee\textsuperscript{3}, S.H. Kim\textsuperscript{3}}
\address{
\textsuperscript{1}Korea Research Institute of Standards and Science, 267 Gajeong-ro, 
Yuseong-gu, Daejeon 34113, Korea\\
\textsuperscript{2}Department of Physics and Astronomy, Seoul National University, 
1 Gwanak-ro, Gwanak-gu, Seoul 08826, Korea\\
\textsuperscript{3}Center for Underground Physics, Institute for Basic Science, 55 Expo-ro, 
Yuseong-gu, Daejeon, 34126, Korea}


\begin{abstract}
Scintillation crystals are commonly used for direct detection of weakly interacting massive particles 
(WIMPs), which are suitable candidates for a particle dark matter. 
It is well known that the scintillation light yields are different for electron recoil and 
nuclear recoil. To calibrate the energies of WIMP-induced nuclear recoil signals, 
the quenching factor (QF) needs to be measured, which is the light yield ratio of the nuclear recoil 
to electron recoil.
Measurements of the QFs for Na and I recoils in a small (2 cm $\times$ 2 cm $\times$ 1.5 cm) 
NaI(Tl) crystal are performed with 2.43-MeV mono-energetic neutrons generated by deuteron-deuteron fusion. 
Depending on the scattering angle of the neutrons, the energies of the recoiled ions vary in the range of
9 - 152 keV 
for Na and 19 - 75 keV for I. The QFs of Na are measured at 9 points with values in the range of 10 - 23 \% 
while those of I are measured at 4 points with values in the  range of  4 - 6 \%.

\end{abstract}

\begin{keyword}
Dark Matter\sep WIMP \sep KIMS \sep NaI(Tl) crystal
\end{keyword}

\end{frontmatter}


\section{Introduction}
Weakly interacting massive particles (WIMPs) have been among the strongest dark matter candidates 
for the past few decades \cite{ref1,ref2}. Several experiments have been designed and performed for the
direct search of WIMPs using various types of detectors \cite{refn1,refn2}. 
Among the various experiments searching for WIMPs, the DAMA/LIBRA group has 
presented very interesting results. 
They demonstrated the detection of an annual modulation effect compatible with a WIMP interaction 
with a high significance of 12.9 $\sigma$, using 250 kg NaI(Tl) scintillation detectors \cite{ref3}. 
However, several other experiments \cite{kims,ref4,ref5,ref6} have not detected positive signals.
Because of the various systematic differences between the experiments, it is difficult 
to draw clear conclusions about the observation by DAMA/LIBRA \cite{ref7}. It is important 
to reproduce the DAMA/LIBRA experiment with the same target material 
using the same or higher sensitivity.

Recently, the Korea Invisible Mass Search (KIMS, at present COSINE-100 
which is a collaborative experiment 
involving KIMS and DM-Ice) started an experiment for the direct search for WIMPs using a NaI(Tl) 
scintillation detector \cite{ref8}, with the same target material as that of DAMA/LIBRA.
The direct detection of WIMPs using a NaI(Tl) scintillation detector is based on the detection of 
the nucleus recoiled by the WIMP-nucleon interaction. The recoiled nucleus loses its kinetic energy 
and a part of the energy is converted into scintillation light. The amount of scintillation 
light can be used to determine the recoil energy of the nucleus. To obtain 
the relation between the nuclear recoil energy and the scintillation light, an energy calibration needs to
be performed.

The energy calibration for nuclear recoil events can be performed using the elastic scattering of energetic neutrons,
various scattering angles, and/or incident energies of neutrons.
The calibration factor $c_{nr}$ can be expressed as a function of the nuclear recoil energy $E_{nr}$ 
and scintillation light $L$ as 

\begin{equation}
c_{nr} = \frac{E_{nr}}{L}.
\label{eq1}
\end{equation}

The energy calibration needs to be repeated for detectors to monitor the stability of $L$,
which is typically performed with gamma sources. 
The calibration factor $c_{er}$ for the gamma calibration can convert the scintillation light 
to the electron recoil equivalent energy $E_{ee}$ as 

\begin{equation}
E_{ee} = c_{er} \times L.
\label{eq2}
\end{equation}

Using Eqs. \ref{eq1} and \ref{eq2}, the nuclear recoil energy can be obtained as 

\begin{equation}
E_{nr} = c_{nr} \times L = c_{nr} \times \frac{E_{ee}}{c_{er}} = QF^{-1} \times E_{ee},
\label{eq3}
\end{equation}

where QF is the quenching factor,
\begin{equation}
QF = \frac{c_{er}}{c_{nr}} = \frac{E_{ee}}{E_{nr}}.
\label{eq4}
\end{equation}

A few research groups, including DAMA, have measured the QFs using radionuclide neutron sources 
with a broad spectrum 
of neutron energies, such as $^{241}$Am-Be or $^{252}$Cf. 
The DAMA group reported constant values of QFs, $QF_{Na}$ = 0.30 $\pm$ 0.01 at the recoil energy range of 6.5 - 97.0 keV 
for Na and $QF_I$ = 0.09 $\pm$ 0.01 at the recoil energy range of 22 - 330 keV for I \cite{ref11}.
Several measurements, using mono-energetic neutrons produced by neutron generators, obtained
consistent results as well \cite{ref12,ref13,ref14,ref15,ref16}. However, certain recent measurements 
on the QF of NaI(Tl) crystals showed significantly different results by systematically considering the threshold 
effects of the efficiencies \cite{ref17,ref18,ref19,newQF}.

We measured the QFs of Na and I using mono-energetic neutrons generated from deuteron-deuteron 
nuclear fusion reaction. The recoil energies of QFs reported here are in the range of 9 - 152 keV 
for Na and 19 - 75 keV for I.

\section{Experiment}

\subsection{Experimental setup}
\label{setup_section}
Mono-energetic neutrons were produced by deuteron-deuteron nuclear fusion reaction using 
a DD109 neutron generator (Adelphi Technology, Inc. \cite{ref22}) at the Korea Research Institute of Standards 
and Science (KRISS). The generator tube was shielded by borated polyethylene (thickness of 40 cm) and 
high-density polyethylene (thickness of 40 cm) successively. The neutrons were extracted 
through a 3.5-cm-diameter hole on the shield. 
This heavy shield fulfills safety regulations. The deuteron beam 
energy was 60 keV. The entire experimental setup was installed at an angle of 90$^{\rm o}$ 
with respect to the deuteron beam.
The neutron energy was measured by a $^{3}$He proportional counter, and the measured neutron energy was 
2.43  $\pm$ 0.03 MeV. 

\begin{figure}
\begin{center}
\includegraphics[width=0.6\textwidth]{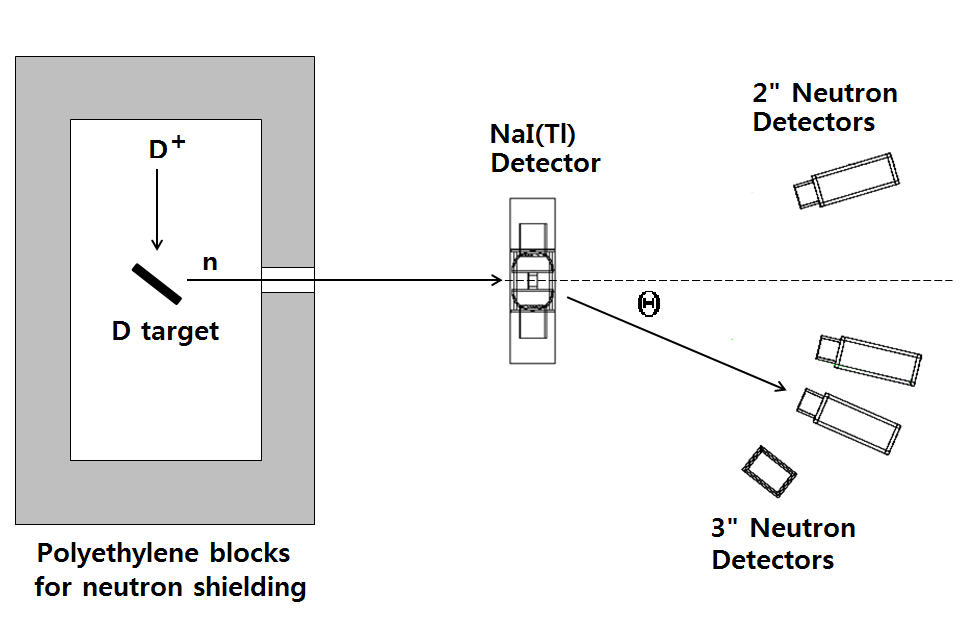}
\caption{Experimental setup for quenching factor (QF) measurement}
\label{Experimental_setup_fig}
\end{center}
\end{figure}

\indent
Figure \ref{Experimental_setup_fig} shows a schematic view of the experimental setup. A NaI(Tl) crystal was 
located at a distance of 150 cm from the target. The size of the crystal was 2 cm $\times$ 2 cm  $\times$ 1.5 cm, 
and the surface of the 2 cm  $\times$ 2 cm side was exposed to the neutron beam.
The typical neutron intensity at the NaI crystal was approximately 490 cm$^{-2}$s$^{-1}$,
which is $\sim$ 2000 s$^{-1}$ on the front face of the crystal.
The NaI(Tl) crystal was the same as one of the crystals used in the COSINE experiment(Crystal-2 in Ref. \cite{ref8}), 
produced by Alpha Spectra, Inc. by the modified Bridgman-Stockbarger method.
The small size of the crystal was chosen to reduce multiple scatterings inside the crystal 
and the spreading angle of the neutrons. Based on a simulation by the GEANT4 toolkit \cite{geant4}, 
the multiple scattering probability was approximately 10 \%. The crystal was encapsulated 
in an aluminum housing with a thickness of 1.52 mm and was coupled to two 3-inch photomultiplier tubes (PMTs) 
with high quantum efficiency (R12669SEL, Hamamatsu Photonics) on two 2 cm $\times$ 1.5 cm sides. 
Quartz blocks with thicknesses of 5 mm were attached between the crystal and the PMTs at both sides to 
achieve the same detector configuration as that of the COSINE-100 experiment.

To tag the neutrons scattered off the Na or I nuclei inside the crystal, BC501A liquid scintillation detectors 
were installed on the plane of the deuteron beam, the deuteron target, and the NaI(Tl) crystal. 
The recoil energy $E_{nr}$ can be expressed by a simple kinetic equation using the incident neutron energy $E_n$,
the scattering angle $\theta$ of the neutron, the mass of the neutron $m_n$,  
and the mass of the recoil nuclide $m_N$ :  
\begin{equation}
E_{nr} = E_n \cdot \{1 + (\frac{m_n cos\theta-\sqrt{{m_N}^2-{m_n}^2 sin^2\theta}}{m_n+m_N})^2\}.
\label{eq5}
\end{equation}

The neutron detectors were installed at 12 different recoil angles from 13$^{\circ}$ 
to 170$^{\circ}$ at distances in the range of 30 - 85 cm from the crystal center. 
The corresponding recoil energies 
were in the range of 6 - 152 keV for Na and 11 - 75 keV for I. Because of the limited space, the measurements 
were performed for three different sets with four different recoil angles. 
Table \ref{NGsim_table} shows the configuration of the three sets of neutron  detectors 
(the size of detectors, distances, and angles), while the corresponding recoil energies for Na and I 
were calculated using Eq. \ref{eq5}. \\

\begin{table}[h]
\begin{center}
\begin{tabular}{c c c c c}
\hline\hline
Set & Size                 & Scattering  & Distance & $E_{nr}$ \\
    & (Diameter $\times$ Length) & angle (degree) & (cm) & (keV)\\
\hline\hline
1 & 5 cm    $\times$ 5 cm    &  13.2 & 82.3 & 5.6 (Na)\\ 
  & 5 cm    $\times$ 5 cm    &  16.4 & 83.6 & 8.7 (Na)\\
  & 5 cm    $\times$ 5 cm    &  26.6 & 84.4 & 22.5 (Na)\\
  & 7.5 cm    $\times$ 9 cm    &  38.2 & 84.0 & 45.2 (Na)\\
\hline
2 & 5 cm    $\times$ 5 cm    &  21.3 & 84.6 & 14.5 (Na)\\
  & 7.5 cm    $\times$ 9 cm    &  59.0 & 46.3 & 101.3 (Na) / 18.7 (I) \\
  & 7.5 cm    $\times$ 9 cm    &  74.7 & 45.0 & 152.1 (Na) / 28.3 (I)\\
  & 7.5 cm    $\times$ 9 cm    & 126.9 & 38.0 & 61.0 (I)\\
\hline
3 & 7.5 cm    $\times$ 9 cm    &  31.0 & 46.3 & 30.3 (Na)\\
  & 7.5 cm    $\times$ 9 cm    &  45.0 & 44.6 & 61.7 (Na) / 11.3 (I) \\
  & 7.5 cm    $\times$ 9 cm    &  51.3 & 52.0 & 78.6 (Na) / 14.4 (I) \\
  & 7.5 cm    $\times$ 9 cm    & 159.4 & 30.7 & 73.7 (I)\\
\hline\hline
\end{tabular}
\caption{Neutron detector configurations for the quenching factor measurements. Because of the 
limited space, the measurements were performed for three different configurations. 
The recoil energies were calculated using Eq. \ref{eq5}.}
\label{NGsim_table}
\end{center}
\end{table}

\subsection{Data acquisition (DAQ) system}
\label{daq_section}
The signals from the NaI(Tl) detector and the neutron detectors were recorded with a sampling rate of 
400 MHz by 10-bit flash analog-to-digital converters(FADCs) from NOTICE, Korea,
with a dynamic range of 1 V \cite{kimsdaq}. Signals from the crystal were 
amplified by 30 times with a custom-made amplifier and sent to the FADC. The additional high-gain 
amplifier for the NaI(Tl) detector enabled  the identification of single photoelectron signals.
Signals from the neutron detectors were sent directly to the FADC.

To prevent PMT noise, a coincidence of signals from PMTs of both sides is required 
within a time window of 200 ns. The first in-coming photoelectron determines the timing of the NaI(Tl) signal. 
The trigger condition for data acquisition requires a time coincidence between the NaI(Tl) detector and one of the four neutron 
detectors. The coincidence time window was 480 ns, which was limited on the front-end DAQ module 
by the embedded software.
For the triggered events, 
the waveforms from the PMTs of the NaI(Tl) detector and the four neutron detectors were recorded by the DAQ system
for a 10 $\mu$s window (2 $\mu$s for the pre-trigger region and 8 $\mu$s for the triggered pulse). 
The event rate was $\sim$ 1.0 Hz. The data were obtained for up to 1,000 recoil events per each recoil 
energy and were recorded for 70, 55, and 25 h for each setup.

\section{Data Analysis}
\subsection{Signal from NaI crystal}
\label{signal_section}
The high-gain, low-noise set of the PMT and the amplifier is capable of providing 
single photoelectron discrimination. To reduce the electrical noise effect and to lower 
the detection threshold, an analysis code was developed for the clustering, which treats 
each local peak as a single photoelectron signal \cite{ref20}. The total charge was calculated 
from the sum of the cluster areas within 1.5 $\mu$s, considering the decay time of the scintillation light of the crystal.
The timing of the signal was determined using the first in-coming cluster.

The energy calibration for the electron equivalent energy was performed with 59.54-keV gamma rays 
from an $^{241}$Am source. 
The linearity of the energy scale at the low-energy region of 1.8 - 22 keV was verified 
with a separate measurement using Compton electrons from165.8-keV gammas from $^{139}$Ce decay. 
The 165.8-keV gammas were scattered by the NaI(Tl) crystal and tagged by LaBr$_{3}$ detectors, 
installed at the various fixed angles.
The energy of Compton electrons inside the NaI(Tl) crystal was measured by the NaI(Tl) detector itself. 
The scattering angle of the gammas was determined by the experimental geometry.
The measured energies of the Compton electrons were compared with the calculated energies 
using the Compton scattering angle of the 165.8-keV gamma ray. 
The values were consistent with each other  within 10 \% \cite{report}.

The photoelectron (p.e.) yield for the small crystal was $\sim$ 14 p.e./keV, which was determined 
by the ratio of the total charge of the 59.54-keV gamma ray to a single p.e. charge.

\subsection{Identification of nuclear recoil events}
\label{neutron_section}
\begin{figure}
\includegraphics[width=1\textwidth]{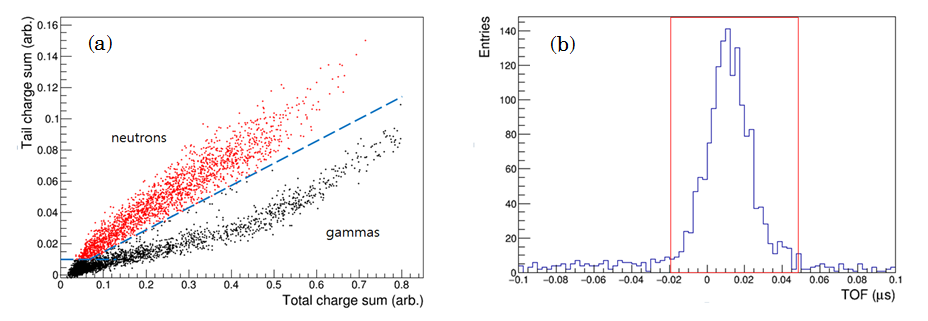}
\caption{(a) PSD for the neutron detector: total charge vs. charge sum 
of the tail section of the neutron detector signal. 
The blue dashed line indicates the selection criteria for neutrons and the red and black dotted points indicate 
neutron and gamma events, respectively. (b) TOF of neutrons from the NaI(Tl) crystal to the neutron detector. 
The events in this spectrum passed the PMT noise cut of the NaI(Tl) detector 
and the electron equivalent energy at the NaI(Tl) detector was higher than 1 keV.
The TOF window for neutron tagging selection is indicated by the red box, which corresponds to 3$\sigma$.}
\label{ND_selection_fig}
\end{figure}

To identify the neutron-induced events in the NaI(Tl) crystal, a coincidence between the NaI(Tl)  
detector and one of the neutron detectors is required.
The neutron detector made of liquid scintillator has an appropriate pulse shape discrimination (PSD) capability to distinguish  
the neutron events from the gamma background. Because neutron-induced events (proton recoil events 
inside the detector) in the liquid scintillator have a longer decay time, 
the PSD against a gamma background was performed using the ratio of the charge sum of the tail section
(50 - 200 ns from the leading edge) to the total charge (over 200 ns). 
Figure \ref{ND_selection_fig}(a) shows the PSD plot for the neutron detector,
where the blue dashed line indicates the cutoff criteria to select neutron-induced events.

The time-of-flight (TOF) of the neutrons scattered off the Na or I nuclei from the NaI(Tl) crystal 
to the neutron detector was constant because the neutrons were monoenergetic. 
For the 2.43-MeV neutrons, the TOFs from the NaI(Tl) crystal to one of the neutron detectors were 
in the range of 14 - 40 ns, calculated using the neutron velocity and the distance 
between the NaI(Tl) crystal and the neutron detector, as shown in Table 1. 
This well-defined TOF enabled the selection of neutron-induced events. 
In the measurement, the neutron TOF was determined by the time difference between the neutron detector 
(BC501A) and the NaI(Tl) detector. The timing of the neutron detector was determined 
by the leading edge of the signal pulse.
The timing of the NaI(Tl) detector was determined by the leading edge of the first cluster of each event.
Figure \ref{ND_selection_fig}(b) shows the neutron TOF spectrum for events, which passed the PMT noise cut
of the NaI(Tl) detector described in section \ref{noisecut_section} and whose electron equivalent energies 
at the NaI(Tl) detector were higher than 1 keV. 
The peak position of the TOF spectrum is not realistic because the time offset was not calibrated. 
From the TOF spectrum, the TOF values for the neutron tagging selection were chosen to be within 3$\sigma$.

\subsection{PMT noise rejection cut for the NaI(Tl) detector}
\label{noisecut_section}
The trigger condition for the NaI(Tl) detector was at least 1 p.e. in each PMT within 200 ns. 
In the low-energy region, PMT noise events were predominantly triggered. 
To eliminate these noise events, we applied two main noise rejection cuts: the charge asymmetry 
between two PMTs, and signal shape discrimination \cite{ref8}.
No correlation between the charge asymmetry and the signal shape discrimination 
was found in the the scatter plot of the two variables defined in Eqs. \ref{eq6} and \ref{eq7}.
The efficiency of the event selection for the charge asymmetry cut was evaluated with the selected events 
applying tighter signal shape discrimination criteria than that for quenching factor analysis, and vice versa.
The efficiency of the event selection of the PMT noise cut was obtained by multiplying two efficiencies.

\begin{figure}
\includegraphics[width=0.5\textwidth]{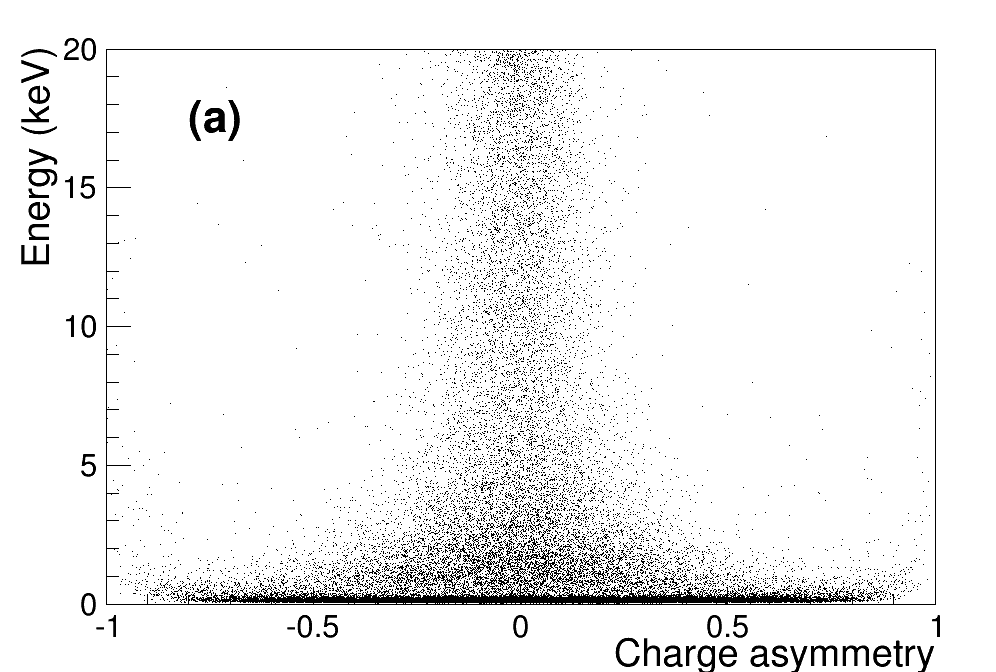}
\includegraphics[width=0.5\textwidth]{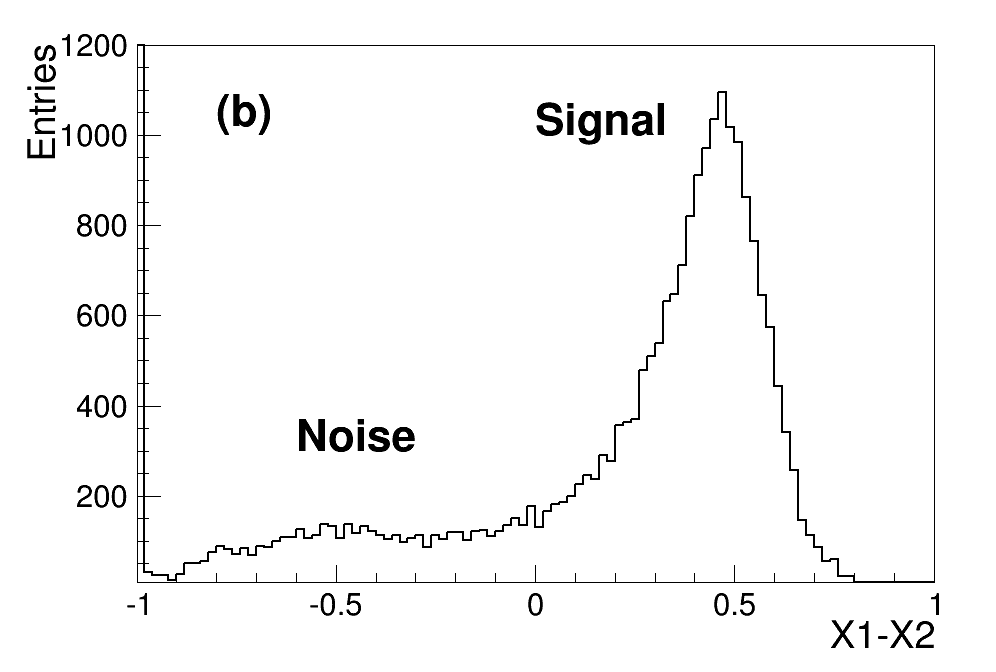}
\caption{(a) Charge asymmetry distribution for neutron recoil events.
Events with large asymmetry values were considered to be noise events. 
(b) X1-X2 distribution for neutron recoil events of 0.25 keV $< E_{ee} <$ 20 keV.
X1 and X2 are defined in Eq. \ref{eq7}. A positive value indicates a high fraction of the slow component, 
which is expected for NaI(Tl) scintillation events. 
A negative value indicates a high fraction of the fast component, which is typical for noise-like events.
}
\label{NaI_selection_fig}
\end{figure}

The PMT noise events typically have a large asymmetry in the total charge of each PMT.
The asymmetry parameter is defined as
\begin{equation}
    Asym=\frac{Q_{PMT1}-Q_{PMT2}}{Q_{PMT1}+Q_{PMT2}},
    \label{eq6}
\end{equation}
where $Q_{PMT}$ denotes the charge sum in each PMT. Figure \ref{NaI_selection_fig}(a) shows 
a scatter plot of the charge asymmetry as a function of measured energy. 
Events with an asymmetry between -0.5 and 0.5 were selected as nominal scintillating events.
The efficiency of the event selection for the charge asymmetry cut was nearly 100 \% above 3 keV and 
became $\sim$ 94 \% at 1 keV.

The signal shape cut is based on the fact that the decay time of the noise pulse
is significantly shorter than that of the typical scintillation signal. 
This was originally developed by the DAMA group 
and they defined ratios of the pulse areas of fast and slow parts \cite{ref21}. 
The fractional charges of slow and fast parts, denoted by X1 and X2, respectively, are defined as
\begin{equation}
    X1=\frac{Q_{100 to 600ns}}{Q_{0 to 600ns}}, X2=\frac{Q_{0 to 50ns}}{Q_{0 to 600ns}},
    \label{eq7}
\end{equation}
where Q is the integrated charge in the time range denoted in the subscript. 
Figure \ref{NaI_selection_fig}(b) shows the distribution of the difference between X1 and X2 (X1-X2). 
Events satisfying 0 $<$ X1-X2 $<$ 0.9 were selected \cite{ref8,ref21}.
The efficiency of the event selection for the signal shape cut was nearly 100 \% above 3 keV and 
became $\sim$ 85 \% at 1 keV.

\begin{figure}
\begin{center}
\includegraphics[width=0.7\textwidth]{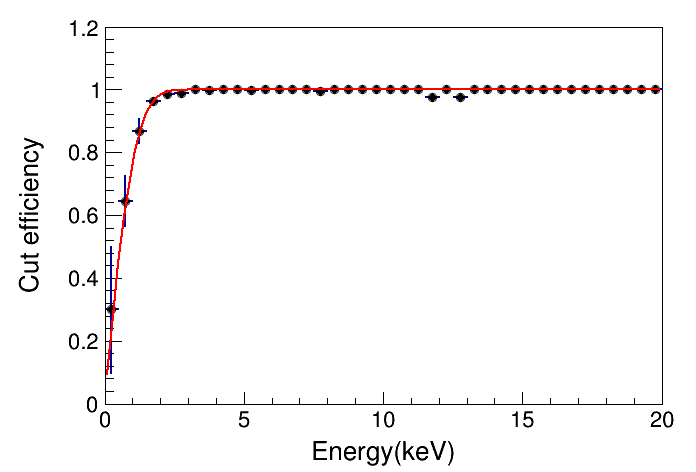}
\caption{Event selection efficiency of PMT noise cut for each 0.5 keV energy bin. 
The red curve is the result of fitting with the error function. The uncertainties of the data points are included in the fit.}
\label{cut_efficiency_fig}
\end{center}
\end{figure}

The effect of the PMT noise cut to the real scintillation signal was analyzed for three sets of the measurements 
described in Table \ref{NGsim_table}, where each set of measurements was independent of other sets. 
The event selection efficiencies of the PMT noise cut for three measurement sets were consistent with each other 
within statistical fluctuations.
The average of three efficiency values was used as the PMT noise cut efficiency for the QF measurement.
The PMT noise cut efficiency as a function of energy  is shown in Figure \ref{cut_efficiency_fig}. 

The cut efficiency ($\epsilon_c$) was fitted with the error function
\begin{equation}
    \epsilon_c(E_{ee}) = p_{c} \times erf(E_{ee} \times q_{c}),
    \label{eq8}
\end{equation}
where $p_c$ is set to unity and $q_c$ is a free parameter.
The result of the fit was $q_c = 0.838 \pm 0.039$ and the reduced chi-square of the fit was 1.2.

\subsection{Determination of trigger efficiency}
\label{trgeff_section}

\begin{figure}
\begin{center}
\includegraphics[width=0.6\textwidth]{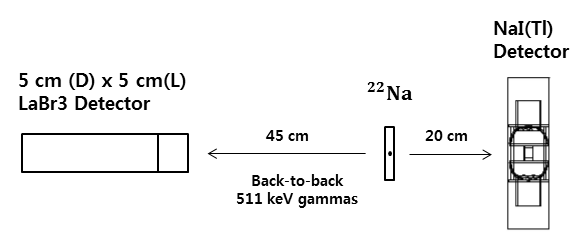}
\caption{Experimental setup to determine trigger efficiency}
\label{trgeff_setup_fig}
\end{center}
\end{figure}

We analyzed the trigger efficiency 
for the low-energy region by performing a separate experiment. We used a $^{22}$Na radioactive source, 
which emits positrons that annihilate into two 511-keV gammas. 
By tagging one of these gammas, we could obtain the response of the NaI(Tl) detector to a 511-keV gamma.
Figure \ref{trgeff_setup_fig} shows a schematic view of the trigger efficiency measurement.
The NaI(Tl) crystal, $^{22}$Na source, and LaBr$_{3}$ detector were installed in one line.
The $^{22}$Na source was covered with a 2-mm-thick copper plate to block the positron emitted 
from the $^{22}$Na decay.
The positron annihilates into two 511-keV gammas, and they fly back-to-back. 
If the NaI(Tl) crystal is hit by a 511-keV gamma, the LaBr$_{3}$ detector is hit by the other 511-keV gamma 
with high probability, and vice versa.

\begin{figure}
\includegraphics[width=1\textwidth]{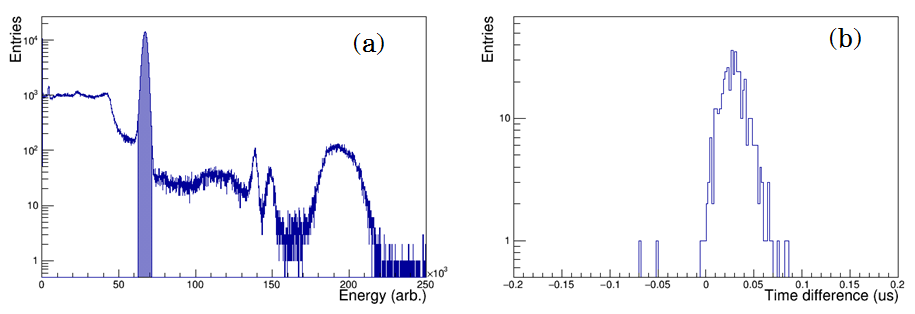}
\caption{Event selection for back-to-back 511 keV gamma-induced events. 
(a) Energy spectrum of LaBr$_{3}$ detector. The blue-filled area indicates the 511 keV peak selected 
for the analysis.
(b) Time difference between the NaI(Tl) detector and LaBr$_{3}$ detector. The offset of the horizontal axis is not calibrated.}
\label{b2b_selection_fig}
\end{figure}

Two independent measurements were carried out. The first measurement was performed with a trigger 
by the LaBr$_{3}$ detector.
The second measurement was performed with a trigger by the NaI(Tl) detector,
which is the same trigger condition as the QF measurement, except for the neutron tagging.
The entire DAQ and all analyses were performed exactly in the same framework for the two measurements.
By comparing the low-energy spectra from the two measurements, we can obtain the trigger efficiency 
of the NaI(Tl) detector.
Figure \ref{b2b_selection_fig}(a) shows the pulse height spectrum of the LaBr$_{3}$ detector. The events 
at the 511-keV peak of the LaBr$_{3}$ data were selected to minimize the background contribution 
in the measurements.
Figure \ref{b2b_selection_fig}(b) shows the distribution of the time differences of the NaI(Tl) detector
and LaBr$_{3}$ detector for the events at the 511-keV peak of the LaBr$_{3}$ detector, 
where the time offset of the horizontal axis is not calibrated.
The time difference distribution shows that the two detectors received hits by back-to-back gammas.

\begin{figure}
\begin{center}
\includegraphics[width=0.7\textwidth]{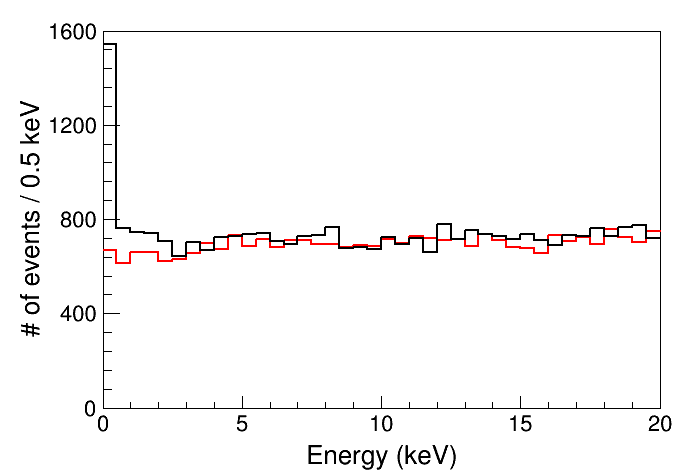}
\caption{Energy spectra of the NaI(Tl) detector.
The red histogram is the spectrum of the NaI(Tl) triggered measurement, 
and the black histogram is that of the LaBr$_{3}$ triggered measurement. 
The first bin of the black histogram has an excess,
which is probably due to the PMT noise events.}
\label{na22_spectrum_fig}
\end{center}
\end{figure}

\begin{figure}
\begin{center}
\includegraphics[width=0.7\textwidth]{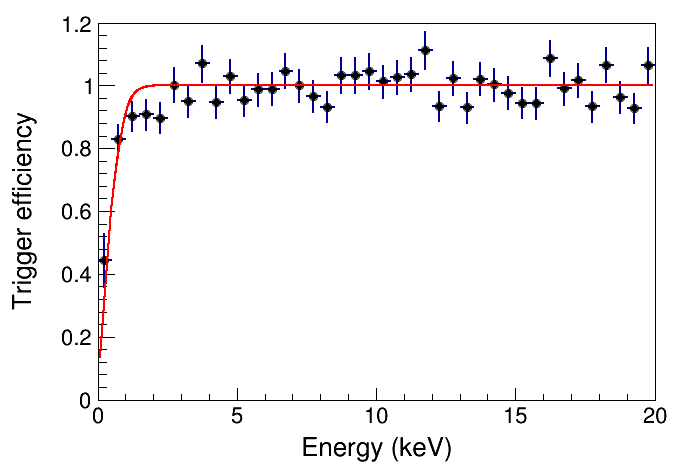}
\caption{Trigger efficiency for each 0.5 keV energy bin. The efficiency was normalized to 1 
for the energy range of 5 - 20 keV. The red curve is the result of fitting with the error function. 
The uncertainties of the data points are included in the fit while the first bin is not included.}
\label{efficiency_fig}
\end{center}
\end{figure}

Asymmetry and signal shape discrimination for the NaI(Tl) detector were applied for the event selection.
For these selected events, the electron equivalent energy of the NaI(Tl) detector
for both measurements is shown in Figure \ref{na22_spectrum_fig}. 
The black histogram corresponds to the first measurement triggered by the LaBr$_{3}$ detector, 
and the red histogram corresponds to the second measurement triggered by the NaI(Tl) detector. 
The energy spectrum for the first measurement shows a large excess in the first bin ($E_{ee} < $ 0.5 keV). 
This excess could not be produced by the Compton scattering of the 511-keV gammas. 
The PMT noise events with energy less than 0.5 keV, apparently survived 
after the PMT noise cut and the coincidence with the neutron detector. 
The random coincidence events of the NaI(Tl) detector with 
those of the LaBr$_{3}$ detector were studied for the LaBr$_{3}$ events above 600 keV.
The energy of the NaI(Tl) detector for those events was mostly below 0.5 keV after all the analysis cuts.
The ratio of the number of surviving events of the NaI(Tl) detector in the first and second measurements 
was considered to be the trigger efficiency of the NaI(Tl) detector, as shown in Figure \ref{efficiency_fig}.
The efficiency above 5 keV was normalized to 1, 
where the trigger efficiency could be assumed to be 100 \%.
In this way, the geometrical efficiency difference between the LaBr$_{3}$ detector and the NaI(Tl) crystal was canceled out 
in all energy regions.
The trigger efficiency ($\epsilon_t$) above 0.5 keV was fitted with the error function
\begin{equation}
    \epsilon_t (E_{ee}) = p_t \times erf(E_{ee} \times q_t),
    \label{eq9}
\end{equation}
where $p_t$ is set to unity and $q_t$ is a free parameter.
The result of the fit is $q_t = 1.20 \pm 0.14$ and its reduced chi-square was 1.1. 
The first bin was not included in the fit 
because the PMT noise events were not completely removed below 0.5 keV.

\section{Results and Conclusion}
The quenching factor can be determined from the ratio of the electron equivalent energy 
to the nuclear recoil energy, as given by Eq. \ref{eq4}.
The nuclear recoil energy can be calculated from simple kinematics using the incident neutron energy and 
the scattering angle in Eq. \ref{eq5}. 
However, considering the detector sizes, the energy spread and the profile of 
the neutron beam results in a  very complicated analytic calculation. 
A Monte Carlo simulation, using GEANT4, version-4.9.6 \cite{geant4}, was performed with realistic geometry, 
including the PMTs and support systems as well as detectors. The neutron beam profile 
at the deuteron target was calculated using the kinematics of the d(d,n)$^{3}$He reaction 
and the deuteron beam profile provided by the manufacturer of the DD109 neutron generator (Adelphi Technology, Inc.). 
The nuclear recoil energy was determined from the deposited energy spectrum of the Na or I recoils
inside the NaI(Tl) crystal in the simulation without the quenching effect. The energy spectrum was fitted 
with the Gaussian distribution and the mean value of the Gaussian distribution was used for the nuclear recoil energy 
of each scattering angle setup. The mean neutron energies obtained from the Gaussian fit are shown 
in Table \ref{Quenching_factor_table},
and the values are consistent with those of the simple calculations (summarized in Table \ref{NGsim_table}) within 1 \%.

\begin{figure}
\begin{center}
\includegraphics[width=1\textwidth]{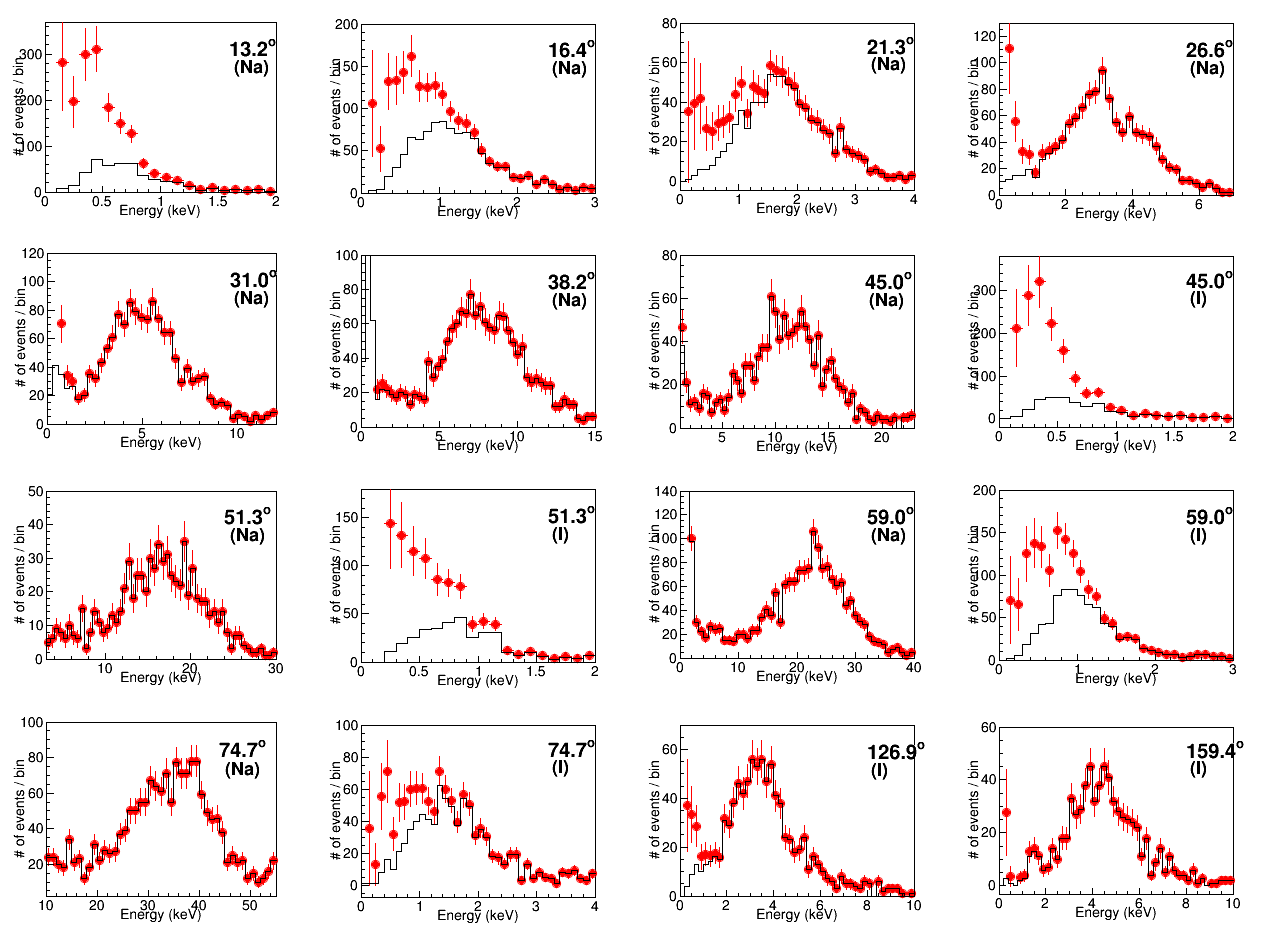}
\caption{Electron equivalent energy spectra for 12 neutron scattering angles. 
The black lines represent the energy spectra before the efficiency correction, 
and the red dots with uncertainties represent those following the application of the efficiency correction 
for the trigger and the analysis cut.
The uncertainties are the quadratic sum of the statistical fluctuation 
and the uncertainty of the efficiency correction.}
\label{depE_fig}
\end{center}
\end{figure}

Figure \ref{depE_fig} shows the measured electron equivalent energy spectra of the nuclear recoil events 
for 12 neutron scattering angles. 
To select the nuclear recoil events, coincidence with the neutron detector is required,
as discussed in section \ref{neutron_section}. 
The PMT noise events were rejected by the PMT noise cut, 
as described in section \ref{noisecut_section}. 
The electron equivalent energy spectra for these selected events are shown by the black histogram in Figure \ref{depE_fig}.
To remove the distortion of the energy spectrum in the low-energy region caused by the trigger and PMT noise cut, 
the trigger efficiency and the cut efficiency were corrected and the resulting spectra are shown in the figures 
by red points with error bars.
The efficiency correction in each energy bin was performed  using
\begin{equation}
N_{corr}(i) = N(i) \cdot {1 \over \epsilon_t} \cdot {1 \over \epsilon_c},
\label{eq10}
\end{equation}
and its uncertainty ($dN_{corr}(i)$) was determined by the quadratic sum of  
the statistical fluctuation of the measurement $dN(i) = \sqrt{N(i)}$ and the uncertainties 
of the efficiency corrections $d\epsilon_c$ and $d\epsilon_t$ as  
\begin{equation}
{dN_{corr}(i) \over N_{corr}(i)} = \sqrt{{dN(i)^2 \over N(i)^2} + {{d\epsilon_c^2 \over \epsilon_c^2 } + {d\epsilon_t^2 \over \epsilon_t^2}}}.
\label{eq13}
\end{equation}
\begin{equation}
d\epsilon_c = {2 \over \sqrt{\pi}} E_{ee} \cdot e^{-(E_{ee} q_c)^2}  dq_c,
\label{eq11}
\end{equation}
\begin{equation}
d\epsilon_t = {2 \over \sqrt{\pi}} E_{ee} \cdot e^{-(E_{ee} q_t)^2}  dq_t,
\label{eq12}
\end{equation}
where $q_c$ and $q_t$ are the fitting  parameters given in Eqs. \ref{eq8} and \ref{eq9}, respectively, 
while $dq_c$ and $dq_t$ are the uncertainties of the fitting parameters, respectively.
Each spectrum in Figure \ref{depE_fig} was fitted with a chi-square fit with a Poisson distribution. 
The fitting range was limited to the energy region above 0.5 keV.

The quenching factors for Na and I were analyzed for 13 points (9 points for Na and 4 for I). 
Three points (1 for Na and 2 for I) were not analyzed 
because the mean of the electron equivalent energy was below 0.5 keV. 
The QFs for Na are in the range of 10 - 23 \% for recoil energies in the range of  9 - 152 keV.
The recoil energy of 9 keV corresponds to an electron equivalent energy of $\sim$ 1 keV, 
which is the expected threshold for the COSINE experiment.
Those for I are in the range of 4 - 6 \% for recoil energies in the range of 19 - 75 keV.
The QFs for Na and I analyzed in this study are summarized in Table \ref{Quenching_factor_table}.

\begin{figure}
\begin{center}
\includegraphics[width=0.6\textwidth]{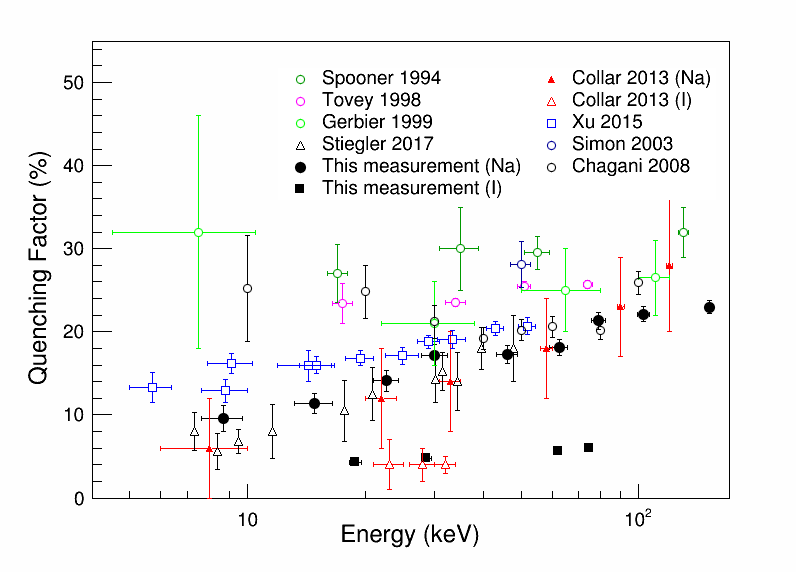}
\caption{QFs for Na and I recoils obtained in this work and their comparison 
with previous measurements. The closed black circles and squares indicate QFs 
in this measurement for Na and I, respectively. For the present measurements, the energy calibration 
for the electron equivalent energy was performed with 59.54-keV gammas from $^{241}$Am.}
\label{Quenching_factor_fig}
\end{center}
\end{figure}

In Figure \ref{Quenching_factor_fig}, the present measurements are compared with previous ones.
The filled circles (Na) and squares (I) correspond to the measurements reported in this study. 
For the QFs for Na, the present measurements are consistent with the recent measurements 
by Collar(red triangles) \cite{ref18}, Xu {\it et. al.}(blue boxes) \cite{ref19}, 
and Stiegler {\it et. al.}(black triangles) \cite{newQF}, 
but the uncertainties in this study are smaller than those of the others. 
For I, the newly measured values are consistent with the results of Collar, but with higher accuracy.

\begin{table}[h]
\begin{center}
\begin{tabular} {c c c c c}
\hline\hline
        & Scattering      & $E_{ee}$  &  $E_{nr}$        & Quenching     \\
Nuclei  & angle (degree)  &  (keV) &  (keV)  & factor ($\%$) \\
\hline\hline
Na & 13.2 &     $<$ 0.5      &   5.8 $\pm$ 1.0 &                \\ 
   & 16.4 &  0.83 $\pm$ 0.07 &   8.7 $\pm$ 1.3 & 9.6  $\pm$ 1.6 \\   
   & 21.3 &  1.68 $\pm$ 0.04 &  14.8 $\pm$ 1.6 & 11.3 $\pm$ 1.2 \\    
   & 26.6 &  3.20 $\pm$ 0.05 &  22.7 $\pm$ 2.0 & 14.1 $\pm$ 1.3 \\ 
   & 31.0 &  5.17 $\pm$ 0.07 &  30.1 $\pm$ 2.2 & 17.2 $\pm$ 1.3 \\ 
   & 38.2 &  7.97 $\pm$ 0.09 &  46.1 $\pm$ 2.8 & 17.3 $\pm$ 1.1 \\ 
   & 45.0 & 11.4 $\pm$ 0.1   &  62.6 $\pm$ 3.2 & 18.1 $\pm$ 0.9 \\ 
   & 51.3 & 16.8 $\pm$ 0.2   &  78.9 $\pm$ 3.6 & 21.3 $\pm$ 1.0 \\ 
   & 59.0 & 22.7 $\pm$ 0.2   & 102.7 $\pm$ 4.1 & 22.1 $\pm$ 0.9 \\  
   & 74.7 & 34.7 $\pm$ 0.3   & 151.6 $\pm$ 5.0 & 22.9 $\pm$ 0.8 \\ 
\hline
 I &  45.0 &  $<$ 0.5        & 11.3 $\pm$ 0.6 &               \\ 
   &  51.3 &  $<$ 0.5        & 14.6 $\pm$ 0.7 &               \\ 
   &  59.0 & 0.80 $\pm$ 0.06 & 18.9 $\pm$ 0.8 & 4.3 $\pm$ 0.4 \\ 
   &  74.7 & 1.35 $\pm$ 0.04 & 28.7 $\pm$ 1.0 & 4.7 $\pm$ 0.2 \\ 
   & 126.9 & 3.47 $\pm$ 0.10 & 62.2 $\pm$ 1.5 & 5.6 $\pm$ 0.2 \\ 
   & 159.4 & 4.44 $\pm$ 0.10 & 74.9 $\pm$ 1.6 & 5.9 $\pm$ 0.2 \\ 
\hline\hline
\end{tabular}
\caption{Summary of quenching factors. The scattering angles were determined by the geometry. 
The electron equivalent energies, $E_{ee}$, were determined 
by the fit for the energy spectrum of the NaI(Tl) detector after applying the correction of 
trigger efficiency and PMT noise cut efficiency. 
The energy calibration was performed with 59.54-keV gammas from $^{241}Am$.  
The nuclear recoil energy was determined by Monte Carlo simulation, as described in the text. 
The quenching factors were calculated using Eq. \ref{eq4}.}
\label{Quenching_factor_table}

\end{center}
\end{table}

\section{Acknowledgements}
This research was supported by the Institute for Basic Science (Korea) under project code IBS-R016-A1. 
H.S.Park and J.H.Kim were supported by the Korea Research Institute of Standards and Science, 
under the project ``Development of measurement standards for radiation''(KRISS-2018-18011053).
S.K.Kim was supported by the grant No. NRF-2016R1A2B3008343.

\section*{References}

\bibliography{mybibfile}

\end{document}